\documentclass[11pt]{article}
\usepackage{amsmath}
\usepackage{amsfonts}
\def\be{\begin{equation}}
\def\ee{\end{equation}}
\def\bea{\begin{eqnarray}}
\def\eea{\end{eqnarray}}
\def\bea*{\begin{eqnarray*}}
\def\eea*{\end{eqnarray*}}
\begin{document}

\title{Spacetimes with all Penrose limits diagonalisable\footnote{I would like to dedicate this paper to the memory of Stephen Siklos, who first derived some of the metrics discussed here. He was a droll and cultured colleague, with the true pedagogic and academic values.}}
\author{Paul Tod\\Mathematical Institute,\\Oxford University}

\maketitle
\begin{abstract}
We consider the problem of finding all space-time metrics for which all plane-wave Penrose limits are diagonalisable plane waves. This requirement leads to a conformally invariant differential condition on the Weyl spinor which we analyse for different algebraic types in the Petrov-Pirani-Penrose classification. The only vacuum examples, apart from actual plane waves which are their own Penrose limit, are some of the nonrotating type D metrics, but some nonvacuum solutions are also identified. The condition requires the Weyl spinor, whenever it is nonzero, to be proportional to a valence-4 Killing spinor with a real function of proportionality.
\end{abstract}

\section{Introduction}
It's well-known that, given a smooth 3-dimensional metric, Riemannian or Lorentzian, coordinates can locally be found in which the metric is diagonal (see \cite{dk} for Riemannian, and \cite{g} for Lorentzian). Call this process \emph{diagonalisation}, then a metric in four or more dimensions cannot always be diagonalised. The problem was considered in \cite{T92} where some non-diagonalisable 4-dimensional Lorentzian metrics were given, and has recently been considered in \cite{gm} where some Riemannian metrics in dimension 4 or more are shown to be non-diagonalisable. In general it is still a difficult problem to decide whether a given 4-metric is diagonalisable, but in \cite{T92} a simple criterion was given for identifying those (Lorentzian, 4-dimensional) plane wave metrics which are diagonalisable. The citerion involves the phase of the remaining complex component of the Weyl tensor, suitably defined. 

In a celebrated paper \cite{P1}, Penrose showed that \emph{every space-time has a plane wave limit}. This plane wave Penrose limit as originally defined, \cite{P1}, entailed defining a coordinate system based on a choice of null-geodesic segment (without conjugate points) in a 4-dimensional space-time and taking a limit defined explicitly in these coordinates, which then leads to a metric in the highly symmetric class of plane waves (see e.g. \cite{es} for an account of these metrics). The limit depends on which null geodesic is chosen so that many limits are possible, some of which may be diagonalisable and some not. 

Some time after Penrose introduced the limit, G\"uven \cite{gu} extended the construction to dimensions greater than 4 and to space-times with additional fields. After this, from about 2002, there was a surge of interest in applications of the Penrose limit to models in string theory. This literature is too large to cite at length, but see for example \cite{b1,NK,pa} and references given in them. In this note the interest is specifically in General Relativity and the case of 4-dimensions, and we exploit the 2-spinor calculus to streamline calculations. 

The question naturally arises of finding conditions on a space-time for all of its Penrose limits to be diagonalisable. It is not the case, as one might naively have hoped, that every diagonalisable space-time has only diagonalisable Penrose limits -- simple counter-examples are provided by the vacuum Kasner metrics (as shown in the Appendix) and by static vacuum metrics which are not type D in the Petrov-Pirani-Penrose classification (see e.g. \cite{pr}), so in particular not the Schwarzschild solution which is in fact both diagonalisable and has all Penrose limits diagonalisable (this follows from the type D case of Proposition 1). In this note, we consider the problem of characterising those space-times which do in fact have all Penrose limits diagonalisable. This turns out to be a strong condition on the Weyl curvature, particularly if the extra condition of vacuum or Einstein is imposed in the space-time, when the only examples are plane waves themselves, the Lobachevski plane waves of Siklos \cite{sik} and some non-rotating type D solutions. In all cases, vacuum or non-vacuum, the condition forces the Weyl spinor to be proportional to a Killing spinor (as defined below) with a real function of proportionality.

The plan of the paper is as follows. We begin in section 2 with a discussion of plane wave metrics and give an account of the Penrose limit which is slightly different from the original but well-adapted for calculation. In section 3 we connect to diagonalisation and obtain the following condition: a space-time has all its Penrose limits diagonalisable iff the following spinor field is zero:
\[\Sigma_{ABCDEA'B'C'D'E'}= i\left(\psi_{(ABCD}\nabla_{E)(A'}\overline{\psi}_{B'C'D'E')}-C.C.\right),\]
where $\psi_{ABCD}$ is the Weyl spinor. This condition has a tensor expression that we give below. The condition is easily seen to be conformally-invariant and indeed any conformally-flat space-time necessarily has all its Penrose limits diagonalisable by the result in \cite{T92}. The analysis of this condition, and in particular its association with Killing spinors, then leads to our main result, Proposition 1. In an appendix we consider Penrose limits of the vacuum Kasner metric.

\medskip

\noindent{\bf{Acknowledgement:}} The work in this article was completed while the author was in residence at the Institut Mittag-Leffler in Djursholm, Sweden as part of the programme `General Relativity, Geometry and Analysis' during September 2019, supported by the Swedish Research Council under grant no. 2016-06596, and I am grateful for the hospitality of the Institute.

\section{Plane waves and the Penrose Limit}
We review some of the theory of plane waves and present a slightly different take on the Penrose limit.
\subsection{The plane wave metric in the Brinkman form}
The reference for this section is \cite{es}. The plane wave metric in the Brinkman form is
\be\label{m1}
g=2du(dv+H(u,\zeta,\overline{\zeta})du)-2d\zeta d\overline{\zeta},\ee
with coordinates $u,v$ real, $\zeta$ complex, and
\be\label{m2}
H=\frac12(\Psi(u)\zeta^2+2\Phi(u)\zeta\overline{\zeta}+\overline{\Psi}(u)\overline{\zeta}^2),\ee
with real $\Phi(u)$ and complex $\Psi(u)$, which we shall see are components of the Ricci and Weyl spinors. We analyse the metric in the Newman-Penrose spin-coefficient formalism (which we assume known; a good reference is \cite{np}): first choose a null tetrad of one-forms
\[L=du,\;\;N=dv+Hdu,\;\;M=-d\zeta,\]
with implied normalised spinor dyad $(o_A,\iota_A)$ and
\[L_a=o_A\overline{o}_{A'} \mbox{  etc.}\]
in the standard way. The corresponding basis of vector fields is
\[D=\partial_v,\;\;\Delta=\partial_u-H\partial_v,\;\;\delta=\partial_{\overline{\zeta}}.\]
The only nonzero spin-coefficient turns out to be
\[\nu=\Psi\zeta+\Phi\overline{\zeta}\]
so that, in particular, the spinor $o_A$ is covariantly constant or parallel,
and then the only nonzero curvature components are
\[\psi_4=\Psi,\;\;\phi_{22}=\Phi.\]
Therefore the curvature spinors can be written out as
\be\label{m3}\psi_{ABCD}=\Psi o_Ao_Bo_Co_D,\;\;\phi_{ABA'B'}=\Phi o_Ao_B\overline{o}_{A'}\overline{o}_{B'},\;\;\Lambda=0.\ee
One calculates at once that
\[\nabla_{A'A}\psi_{BCDE}=\dot{\Psi}\overline{o}_{A'}o_Ao_Bo_Co_Do_E.\]
Below we shall be interested in the Hermitian spinor field
\be\label{m4}
\Sigma_{abcde}=\Sigma_{ABCDEA'B'C'D'E'}:=i(\psi_{(ABCD}\nabla_{E)(A'}\overline{\psi}_{B'C'D'E')}-C.C.).\ee
In tensors, $\Sigma_{abcde}$ is a constant multiple of the trace-free part (on all indices) of the tensor
\[C_{(a\;\;b}^{*\;p\;\;q}\nabla_cC_{d|p|e)q}-C_{(a\;\;b}^{\;\;p\;\;q}\nabla_cC^*_{d|p|e)q},\]
where, as usual, the vertical bars isolate indices which should be omitted from the symmetrisation.

For the plane waves currently considered, (\ref{m4}) reduces to
\be\label{m5}
\Sigma_{ABCDEA'B'C'D'E'}=i(\Psi\dot{\overline{\Psi}}-\overline{\Psi}\dot{\Psi})o_Ao_Bo_Co_D
\overline{o}_{A'}\overline{o}_{B'}\overline{o}_{C'}\overline{o}_{D}\overline{o}_{E'},\ee
so that there is essentially just one non-trivial component, and it was shown in \cite{T92} that the metric (\ref{m1}) can be diagonalised by choice of coordinates iff this component vanishes, therefore iff $\Sigma_{abcde}$ as in (\ref{m5}) vanishes.

The plane wave metric generically has five linearly independent Killing vectors: one is $K=\partial_v$ and the other four are harder to see in this metric form but easier to see in the Rosen form (again, see e.g. \cite{es} for this). The metric also admits a scaling invariance: for constant, complex $\lambda$, consider the transformation
\be\label{s1}(u,v,\zeta,\Psi,\Phi)\rightarrow(\hat{u},\hat{v},\hat{\zeta},\widehat{\Psi},\widehat{\Phi})=((\lambda\overline{\lambda})^{-1}u,
\lambda\overline{\lambda}v,(\overline{\lambda}/\lambda)\zeta,\lambda^4\Psi,(\lambda\overline{\lambda})^2\Phi),\ee
then the metric is unchanged. We regard two plane waves as equivalent if
\[\hat{\Psi}(\hat{u})=\lambda^4\Psi(u),\;\; \hat{\Phi}(\hat{u})=(\lambda\overline{\lambda})^2\Phi(u),\]
since then they are related by this scaling.

We will consider the null geodesics of this metric. The Killing vector $K$ gives a conserved quantity $E:=K_aV^a=du/ds$ where $V^a$ is the null tangent to a geodesic and $s$ is a choice of affine parameter. If $E$ is zero, then the null geodesic has tangent $K$ and we can choose the spinor $o^A$ as parallelly-propagated tangent to it -- this 3-dimensional class of null geodesics (call it the first class) are parallel to the repeated Principal Null Direction of the Weyl spinor. If $du/ds\neq 0$ (call this the second class) then we may set $E$ to be one by choice of affine parameter, and take the spinor field $\alpha^A=\iota^A+\omega o^A$ to be parallelly-propagated along the null geodesic, which entails
\[0=\alpha^B\overline{\alpha}^{B'}\nabla_{BB'}\alpha^A=(\omega\overline{\omega}D+\omega\delta+\overline{\omega}\overline{\delta}+\Delta)(\iota^A+\omega o^A)\]
\[=(\nu+\frac{d\omega}{ds})o^A,\]
and so $d\omega/ds=-\nu$. The tangent vector is
\[V=\omega\overline{\omega}D+\omega\delta+\overline{\omega}\overline{\delta}+\Delta\]
\[=\omega\overline{\omega}\partial_v+\omega\partial_{\overline{\zeta}}+\overline{\omega}\partial_\zeta+(\partial_u-H\partial_v)\]
from which one reads off
\[du/ds=1,\;\;d\zeta/ds=\overline{\omega},\;\;dv/ds=\omega\overline{\omega}-H.\]
Along a geodesic in the first class we note that the contractions of the curvature spinors with the tangent spinor, namely $\psi_{ABCD}o^Ao^Bo^Co^D$ and $\phi_{ABA'B'}o^Ao^B\overline{o}^{A'}\overline{o}^{B'}$, are both zero. For any geodesic in the second class, which is 5-dimensional, we have instead
\[\psi_{ABCD}\alpha^A\alpha^B\alpha^C\alpha^D=\Psi,\;\;\phi_{ABA'B'}\alpha^A\alpha^B\overline{\alpha}^{A'}\overline{\alpha}^{B'}=\Phi\]
i.e. the same functions for every geodesic in this class. The scaling transformation is relevant here: if we choose the spinor $\hat{\alpha}^A=\lambda\alpha^A$ as the parallelly-propagated tangent to the null geodesic, for constant, complex $\lambda$ then $du/d\hat{s}=\lambda\overline{\lambda}$ and $\Psi,\Phi$ are replaced by $\hat{\Psi},\hat{\Phi}$ as in (\ref{s1}).

Now the Penrose limit \cite{P1} can be obtained as follows: given any null geodesic $\Gamma$ in any space-time $M$, choose a parallelly-propagated spinor $\alpha^A$ tangent to $\Gamma$, determining the affine parameter $u$ up to additive constant by $\alpha^A\overline{\alpha}^{A'}\nabla_{AA'}u=1$, and calculate
\[\Psi(u)=\psi_{ABCD}\alpha^A\alpha^B\alpha^C\alpha^D,\;\;\Phi(u)=\phi_{ABA'B'}\alpha^A\alpha^B\overline{\alpha}^{A'}\overline{\alpha}^{B'}.\]
Then putting these in (\ref{m1}) determines a plane-wave metric from $(M,\Gamma)$, and a different constant rescaling of $\alpha^A$ gives an equivalent plane-wave metric. Thus given a null geodesic in $M$, one has constructed a plane wave space-time in which an open 5-dimensional subset of null geodesics have equivalent data $(\Psi,\Phi)$. It's not hard to show this is equivalent to Penrose's construction \cite{P1}, and it is often computationally simpler.

\section{The connection to diagonalisability}
If the result of taking the Penrose limit of $(M,\Gamma)$ is a diagonalisable plane wave then, by a result in \cite{T92}, we must have
\[\Psi\dot{\overline{\Psi}}-\overline{\Psi}\dot{\Psi}=0\]
along $\Gamma$. Therefore, if every Penrose limit of a given space-time $M$ is diagonalisable then we must have this condition holding at every point and in every null direction so that
\be\label{c1}
\Sigma_{ABCDEA'B'C'D'E'}:=i\left(\psi_{(ABCD}\nabla_{E)(A'}\overline{\psi}_{B'C'D'E')}-C.C.\right)=0.\ee
This is the condition that we wish to analyse. We first recall the definitions of the two scalar invariants $I,J$ of the Weyl spinor:
\[I:=\psi_{ABCD}\psi^{ABCD},\;\;\;J:=\psi_{ABCD}\psi_{PQ}^{\;\;\;\;\;\;AB}\psi^{CDPQ},\] 
and then recall the definition of a (valence-$n$) Killing spinor from \cite{pw} as a symmetric spinor field $\omega_{A_1\cdots A_n}$ satisfying
\[\nabla_{A'(A}\omega_{A_1\cdots A_n)}=0.\]

Then we organise the results according to the algebraic type of the Weyl spinor (in the Petrov-Pirani-Penrose classification, see e.g. \cite{pr}) and summarise them in a Proposition:

\medskip

{\bf{Proposition 1}}

\medskip

Given a space-time $M$ in which (\ref{c1}) holds, the Weyl spinor is always proportional to a valence-4 Killing spinor with a real function of proportionality. Conversely, if the Weyl spinor is proportional to a valence-4 Killing spinor with a real function of proportionality then (\ref{c1}) holds. As regards examples, if the Weyl spinor is:
\begin{enumerate}
\item{\bf{zero}} then all Penrose limits are diagonalisable.
\item{\bf{type N}} then the Weyl spinor is proportional to the fourth power of a valence-1 Killing spinor with a real function of proportionality, all Penrose limits are diagonalisable but the only vacuum examples are already plane waves and the only Einstein examples are Lobachevski plane waves (see \cite{sik}). The nonvacuum examples are easy to find.
\item{\bf{type $(3,1)$ or $(2,1,1)$}} then there are no vacuum or Einstein examples (there may be nonvacuum examples).
\item{\bf{type $D$}} then the vacuum examples must have $\psi_2$, the only nonzero component of the Weyl spinor, real. These conditions lead to a short list of examples, which includes the Schwarzschild metric. For vacuum or non-vacuum examples the Weyl spinor is proportional to the square of a valence-2 Killing spinor with a real function of proportionality.
\item{\bf{algebraically general}} then there are no vacuum or Einstein examples although the Weyl spinor is proportional to an indecomposable valence-4 Killing spinor with a real function of proportionality, and some nonvacuum examples, related to Kobak's \emph{doubly Hermitian} metrics, \cite{kob}, are easy to find.
\end{enumerate}

\noindent{\bf{Proof}}

The converse in the second sentence is easy to establish: if $\omega_{ABCD}=F\psi_{ABCD}$ is Killing spinor with real $F$ then
\[0=\nabla_{A'(A}\omega_{BCDE)}=F(\nabla_{A'(A}\psi_{BCDE)}+F^{-1}(\nabla_{A'(A}F)\psi_{BCDE)}),\]
when (\ref{c1}) clearly holds.

That the Weyl spinor is proportional to a valence-4 Killing spinor is established type by type. In the order of the proposition, suppose that the Weyl spinor is:
\begin{enumerate}
\item {\bf{zero}} so the metric is conformally flat, and (\ref{c1}) is vacuously satisfied, $\Psi(s)$ vanishes for every null geodesic and all Penrose limits of $M$ are diagonalisable (by the result in \cite{T92}).
\item {\bf{type N}}, so that for some spinor field $o_A$ the Weyl spinor takes the form $\psi_{ABCD}=o_Ao_Bo_Co_D$, then (\ref{c1}) can be written
\be\label{c11}o_{(A}o_Bo_C\chi_{DE)}^{\;\;\;\;\;\;(A'B'}\overline{o}^{C'}\overline{o}^{D'}\overline{o}^{E')}=C.C.\ee
where
\[\chi_{AB}^{\;\;\;\;\;\;A'B'}=o_{(A}\nabla_{B)}^{\;\;(A'}\overline{o}^{B')}.\]
Now (\ref{c11}) forces reality of $\chi_{ABA'B'}$, when we deduce that
\be\label{m99}\nabla_{AA'}o_B=\epsilon_{AB}\rho_{A'}+V_{AA'}o_B,\ee
for some spinor $\rho_{A'}$ and real vector $V_{AA'}=V_a$. Contract this with $\nabla_C^{\;\;A'}$:
\[\nabla_C^{\;\;A'}\nabla_{AA'}o_B=\epsilon_{AB}\nabla_C^{\;\;A'}\rho_{A'}+\nabla_C^{\;\;A'}\left(V_{AA'}o_B\right)\]
\[=\epsilon_{AB}\nabla_C^{\;\;A'}\rho_{A'}+o_B\nabla_C^{\;\;A'}V_{AA'}+\epsilon_{CB}V_{AA'}\rho^{A'}+\frac12\epsilon_{AC}(V_eV^e)o_B,\]
using (\ref{m99}) again. Now symmetrise over $CAB$: on the left we obtain
\[=-\psi_{ABCD}o^D=0,\]
by the form of $\psi_{ABCD}$, and on the right all terms vanish except that with the derivative of $V_a$, to leave
\[o_{(B}\nabla_A^{\;\;A'}V_{C)A'}=0,\]
whence
\[\nabla_{(A}^{\;\;A'}V_{C)A'}=0,\]
and since $V_a$ is real, we have
\[\nabla_{[a}V_{b]}=0,\]
so the 1-form $V_a$ is closed, therefore (locally) exact and $V_a=\nabla_aV$ for some real function $V$. Set $\omega_A=e^{-V}o_A$ to find
\be\label{c12}
\nabla_{AA'}\omega_B=i\epsilon_{AB}\pi_{A'},\ee
for a spinor field $\pi_{A'}$ proportional to $\rho_{A'}$. This is the \emph{twistor equation}, \cite{p0}, the pair $(\omega^A,\pi_{A'})$ defines a twistor and $\omega^A$ itself is a valence-1 Killing spinor. Also
\be\label{c13}\psi_{ABCD}=e^{4V}\omega_A\omega_B\omega_C\omega_D,\ee
so that the Weyl spinor is proportional to a valence-4 Killing spinor (since a symmetrised outer product of Killing spinors is evidently a Killing spinor), with a real function of proportionality, $e^{4V}$.

Space-times admitting a solution of (\ref{c12}) have been classified in \cite{L} (though equation (\ref{c13}) imposes an extra condition on them). We may summarise the classification as
\begin{enumerate}
\item those for which the vector field $\omega^A\overline{\omega}^{A'}$, which is a conformal Killing vector, is twisting (equivalently, the twistor $(\omega^A,\pi_{A'})$ is non-null); these were shown in \cite{LN} to be Fefferman metrics of 3-dimensional CR-structures. They cannot be Einstein or vacuum (except trivially i.e. when flat).
\item those for which it is non-twisting but with $\pi_{A'}\neq 0$; these were given explicitly in \cite{L} and include the next two classes.
\begin{enumerate}
\item the only Einstein examples have $\pi_{A'}$ nonzero and proportional to $\overline{\omega}_{A'}$; these are the Lobatchevski plane waves of Siklos \cite{sik}.
\item those for which $\pi_{A'}=0$; these are pp-waves, which we'll discuss shortly. They can be vacuum but not Einstein.
\end{enumerate}
\end{enumerate}

The pp-wave metric can be taken to be (\ref{m1}) but with a general (real, smooth) function $H(u,\zeta,\overline{\zeta})$. Imposing (\ref{c1}) on a pp-wave requires $H_{\zeta\zeta}/H_{\overline{\zeta}\,\overline{\zeta}}$ to be constant. By constant phase change of $\zeta$ we can take this constant to be one and then, if we set $\zeta=x+iy$, the condition reduces to $H_{xy}=0$ and is solved by
\[H=f(u,x)+g(u,y).\]
These give the family of pp-waves all of whose Penrose limits are diagonalisable. The only vacuum metrics among them have $H$ harmonic in $x,y$ when they are in fact plane waves.


\item{\bf{types $(3,1)$ and $(2,1,1)$}} First for $(3,1)$ we may take
\[\psi_{ABCD}=\psi o_{(A}o_Bo_C\iota_{D)}\]
in terms of spinor fields $o_A,\iota_A$ which we may assume normalised by $o_A\iota^A=1$, and a complex function $\psi$. If we restrict to vacuum then $o^A$ is geodesic and shear-free (which we henceforth abbreviate as gsf) but if instead we suppose just that $\Sigma$ vanishes then
\[0=o^Ao^Bo^C\Sigma_{ABCD}^{\;\;\;\;\;\;\;\;\;\;A'B'C'D'}=-io^Ao^Bo^C\overline{\psi}^{(A'B'C'D'}\nabla^{E')}_{\;\;(A}\psi_{BCDE)}.\]
The factor in $\overline{\psi}_{A'B'C'D'}$ is irrelevant and can be omitted, along with the factor $-i$, leaving
\[0=o^Ao^Bo^C\nabla^{E'}_{\;\;(A}(\psi o_Bo_Co_D\iota_{E)})=\psi o_Do_Eo^Bo^C\nabla^{E'}_{\;\;B}o_C,\]
so that $o^A$ is gsf automatically, from the vanishing of $\Sigma$. Next calculate
\[0=\iota^A\iota^B\iota^C\iota^D\iota^E\Sigma_{ABCD}^{\;\;\;\;\;\;\;\;\;\;A'B'C'D'}
=-i\iota^A\iota^B\iota^C\iota^D\iota^E\overline{\psi}^{(A'B'C'D'}\nabla^{E')}_{\;\;(A}\psi_{BCDE)}.\]
Again drop irrelevant factors to find
\[0=\iota^A\iota^B\iota^C\iota^D\iota^E\nabla^{E'}_{\;\;A}\left(\psi o_Bo_Co_D\iota_E\right)\]
whence $\iota^A$ is also gsf. This is sufficient to show that there are no vacuum or Einstein solutions like this as a consequence of the Goldberg-Sachs Theorem \cite{es} - recall this states that, for vacuum or Einstein metrics in dimension 4, a spinor field is gsf iff it is a repeated Principal Null Direction (or PND) of the Weyl spinor, and $\iota_A$ is not a \emph{repeated} PND, though it is a PND. The Goldberg-Sachs Theorem also rules out certain classes of nonvacuum solutions (for example, Einstein-Maxwell solutions with aligned Maxwell fields), but there may be others. We shall see next that (\ref{c1}) forces the Weyl spinor to be proportional to a Killing spinor with a real function of proportionality. This is a little messy as both scalar invariants, $I$ and $J$, vanish in this case.

From (\ref{c1}) then with $\psi_{ABCD}=\psi o_{(A}o_Bo_C\iota_{D)}$, first contract with $\iota^A\iota^B\iota^C\iota^D$ to find
\be\label{31}\frac{1}{5}\psi\iota_E\iota^D\nabla_{D(A'}\overline{\psi}_{B'C'D'E')}=\overline{\psi}_{(A'B'C'D'}W_{E')E},\ee
with
\[W_{EE'}=\iota^A\iota^B\iota^C\iota^D\nabla_{E'(A}\psi_{BCDE)}.\]
Since, as we saw above, $\iota^A$ is gsf, necessarily $\iota^EW_{EE'}=0$ and $W_{EE'}=\iota_E\eta_{E'}$ for some $\eta_{E'}$. Now we may cancel $\iota_E$ from (\ref{31}) to leave
\be\label{32}\iota^D\nabla_{D(A'}\overline{\psi}_{B'C'D'E')}=\overline{\psi}_{(A'B'C'D'}\eta_{E')},\ee
where we've redefined $\eta$ to absorb the factor $\psi/5$. Next contract (\ref{c1}) with $o^Ao^B$ and use
\[o^Ao^B\nabla_{E'(A}\psi_{BCDE)}=o_Co_Do_E\xi_{E'}\]
for some $\xi_{E'}$ which follows since $o^A$ is gsf. We obtain
\[-\psi o_Co_Do_Eo^A\nabla_{A(A'}\overline{\psi}_{B'C'D'E')}=o^Ao^B\overline{\psi}_{(A'B'C'D'}\nabla_{E')(A}\psi_{BCDE)}\]
\[=o_Co_Do_E\overline{\psi}_{(A'B'C'D'}\xi_{E')}.\]
Cancel $o_Co_Do_E$ and absorb the factor $-\psi$ into $\xi_{E'}$ to obtain
\be\label{33}o^A\nabla_{A(A'}\overline{\psi}_{B'C'D'E')}=\overline{\psi}_{(A'B'C'D'}\xi_{E')}.\ee
Now put (\ref{32}) and (\ref{33}) together to deduce that
\[\nabla_{A(A'}\overline{\psi}_{B'C'D'E')}=W_{A(A'}\overline{\psi}_{B'C'D'E')},\]
for some vector $W_{AA'}$ (to be explicit, $W_{AA'}=o_A\eta_{A'}-\iota_A\xi_{A'}$). Impose (\ref{c1}) to deduce that $W_{AA'}$ is real, so that, taking the complex conjugate,
\[\nabla_{A'(A}\psi_{BCDE)}=W_{A'(A}\psi_{BCDE)},\]
and apply $\nabla_F^{\;\;A'}$, symmetrising over $ABCDEF$ to deduce that $W_a$ is a gradient, say $W_a=\nabla_aW$. We've found that
\[\omega_{ABCD}:=e^{-W}\psi_{ABCD}\]
is a valence-4 Killing spinor, proportional to the Weyl spinor with a real function of proportionality (namely $e^{-W}$).

\medskip

Next for type $(2,1,1)$ we can dispose of the vacuum case by the same argument as for type $(3,1)$ here: by (\ref{c1}) any PND of the Weyl spinor is gsf but in vacuum only repeated ones should be, so there are no vacuum solutions of this type, but there could be nonvacuum solutions. It will follow from the argument in the algebraically general case that, since the scalar invariant $I$ is nonzero in this case the Weyl spinor is proportional to a Killing spinor with a real function of proportionality.

\item {\bf{type D}} We recall that vacuum type D solutions always admit a valence-2 Killing spinor $\omega_{AB}$ \cite{pw} and the Weyl spinor is related to the Killing spinor and a normalised spinor dyad $(o_A,\iota_A)$ by
\[\psi_{ABCD}=6\psi_2o_{(A}o_{B}\iota_C\iota_{D)}=\psi_2^{5/3}\omega_{(AB}\omega_{CD)},\]
absorbing a constant numerical factor into the definition of $\omega_{AB}$. The valence-2 Killing spinor satisfies
\[\nabla_{A'(A}\omega_{BC)}=0\]
so that
\[\Sigma_{ABCDEA'B'C'D'E'}=\omega_{(AB}\omega_{CD}W_{E)(A'}\overline{\omega}_{B'C'}\overline{\omega}_{D'E')},\]
with
\[W_e=i|\psi_2|^{2/3}(\overline{\psi}_2\nabla_e\psi_2-\psi_2\nabla_e\overline{\psi}_2).\]
Thus (\ref{c1}) holds in the vacuum case iff $\psi_2/\overline{\psi}_2$ is constant and then the Weyl spinor is proportional to the valence-4 Killing spinor $\omega_{(AB}\omega_{CD)}$ with a real function of proportionality. Since type D vacuum solutions have been classified completely \cite{kin} we can read off the ones with constant $\psi_2/\overline{\psi}_2$. These are the Schwarzschild-like solutions, the static C-metric and Kinnersley's Case $IV A$ restricted by $a=0$ (this example is exceptional: $\psi_2$ isn't real but a constant multiple of it is, which is sufficient for (\ref{c1})) and his Case $IV B$.

Similarly \cite{hpsw} showed that the charged Kerr metric admits a Killing spinor defined in the same way but now $\psi_2$ is complex unless $a=0$, so only in this case (i.e. Reissner-Nordstr\"om) does (\ref{c1}) hold. There may be other nonvacuum solutions, for all of which the Weyl spinor will be proportional to a valence-4 Killing spinor with a real function of proportionality by the argument in the next section, since the scalar invariants $I,J$ are nonzero. Also, since the valence-4 Killing spinor, say $\omega_{ABCD}$ is proportional to the Weyl spinor, there must in fact be a valence-2 Killing spinor $\omega_{AB}$ with $\omega_{ABCD}=\omega_{(AB}\omega_{CD)}$. However one type D nonvacuum solution which fails is the G\"odel metric. This is known to admit a valence-2 Killing spinor with the Weyl spinor proportional to its square (\cite{kal}, \cite{ross}) but $\psi_2$ is not real so that (\ref{c1}) does not hold.

\item {\bf{algebraically general}}
Now we can suppose that at least one of the scalar invariants $I,J$ is nonzero (as we also could in type $(2,1,1)$ and type D). We explore some general consequences of (\ref{c1}); write it
\be\label{c2}\psi_{(ABCD}\nabla_{E)(A'}\overline{\psi}_{B'C'D'E')}=\overline{\psi}_{(A'B'C'D'}\nabla_{E')(A}\psi_{BCDE)}\ee
and contract with $\overline{\psi}_{A'B'C'D'}$ to obtain
\[\psi_{(ABCD}W_{E)E'}=\frac{3}{5}\overline{I}\nabla_{E'(A}\psi_{BCDE)}\]
for some vector $W_{EE'}$, possibly complex, and with $I=\psi_{ABCD}\psi^{ABCD}$. Assume $I\neq 0$ and set $U_a=\frac{5}{3\overline{I}}W_a$, then substitute back into (\ref{c2}) to obtain
\[\psi_{(ABCD}(U_{E)(A'}-\overline{U}_{E)(A'})\overline{\psi}_{B'C'D'E')}=0,\]
which forces $U_a-\overline{U}_a$ to vanish. Therefore
\be\label{c3}\nabla_{E'(A}\psi_{BCDE)}=U_{E'(A}\psi_{BCDE)},\ee
with a real vector $U_e$. Contract this with $\psi^{ABCD}$ to obtain on the right
\[\psi^{ABCD}U_{E'(A}\psi_{BCDE)}=\frac{3}{5}IU_{EE'}\]
and on the left
\[\psi^{ABCD}\nabla_{E'(A}\psi_{BCDE)}=\frac{1}{5}\psi^{ABCD}\nabla_{EE'}\psi_{ABCD}+\frac{4}{5}\psi^{ABCD}\nabla_{E'A}\psi_{BCDE}\]
\[=\frac{1}{10}\nabla_{EE'}I+\frac{4}{5}\left(\nabla_{E'A}(\psi^{ABCD}\psi_{BCDE})-\psi_{BCDE}\nabla_{E'A}\psi^{ABCD}\right).\]
If we restrict to vacuum then the vacuum Bianchi identity gives the vanishing of the term $\nabla_{E'A}\psi^{ABCD}$ and this simplifies to
\[=\frac{1}{10}\nabla_{EE'}I+\frac{4}{10}\nabla_{EE'}I=\frac12\nabla_{EE'}I,\]
so that, since we're assuming $I\neq 0$ we find
\[U_a=\frac{5}{6}\nabla_aI/I.\]
In particular this must be real so that necessarily $I/\overline{I}$ must be constant.

Define
\[\omega_{ABCD}=I^k\psi_{ABCD}\]
for $k$ to be chosen, then (\ref{c1}) entails
\[\nabla_{A'A}\omega_{BCDE}=I^k(U_{A'(A}\psi_{BCDE)}+kI^{-1}I_{A'A}\psi_{BCDE})\]
so if we choose $k=-5/6$ then
\[\nabla_{A'(A}\omega_{BCDE)}=0,\]
and $\omega_{ABCD}$ is a valence-4 Killing spinor. It is known \cite{tdc}, \cite{j} that, wnenever one is present, a valence-4 Killing spinor is always proportional to the Weyl spinor unless this is zero.

If $I=0$ we can use $J$: contracting (\ref{c3}) with $\psi_{PQ}^{\;\;\;\;\;\;AB}\psi^{CDPQ}$ and still restricting to vacuum, by a similar calculation we obtain instead
\[U_a=\frac{5}{9}\nabla_aJ/J,\]
where $J=\psi_{ABCD}\psi_{PQ}^{\;\;\;\;\;\;AB}\psi^{CDPQ}$, assuming this is not zero. Thus necessarily $J$ is real and, if $IJ\neq 0$, then from the two expressions for $U_a$ we conclude that $I^3/J^2$ must be constant. If $I=0$ we define the valence-4 Killing spinor as
\[\omega_{ABCD}=J^{-5/9}\psi_{ABCD}.\]

There cannot be vacuum solutions in this class: it was observed in \cite{j} (and also in \cite{tdc}, and was very probably known to Penrose) that any principal spinor of a Killing spinor is gsf. So if a space-time has an algebraically general Weyl spinor and admits a valence-4 Killing spinor then there are four linearly independent gsf spinors, which contradicts the Goldberg-Sachs Theorem for vacuum or Einstein.

However, there are non-vacuum Hermitian (so Riemannian) examples with four linearly independent gsf congruences (see \cite{kob}) and some of these will have real Lorentzian (Wick-rotated) sections. We'll show in the next subsection that they do have valence-4 Killing spinors, but end the proof of the Proposition here.

\end{enumerate}
\subsection{Kobak's metrics}
In \cite{kob} the metric form
\[g=dzd\overline{z}+fdwd\overline{w}\]
is considered with smooth real $f$. This is evidently Hermitian for the complex structure $I_1$ with $(0,1)$-forms $d\overline{z},d\overline{w}$. One defines a second complex structure $I_2$ with $(0,1)$-forms
\[d\overline{z}+fdw,\;\;\;d\overline{w}-dz,\]
which can be checked to be  integrable, given an equation on $f$, and also Hermitian.

(To see this take the orthonormal $(0,1)$ forms for $I_1$ to be
\[\Theta^{\overline{1}}=d\overline{z},\;\;\;\Theta^{\overline{2}}=f^{1/2}d\overline{w},\]
while for $I_2$ choose
\[\Phi^{\overline{1}}=\frac{1}{(1+f)^{1/2}}(\Theta^{\overline{1}}+f^{1/2}\Theta^2),\;\;
\Phi^{\overline{2}}=\frac{1}{(1+f)^{1/2}}(f^{-1/2}\Theta^{\overline{2}}-\Theta^1),
\]
and note that the transformation between bases is Hermitian.)

Integrability requires
\[f_z-f_{\overline{w}}=0\mbox{   so that   }f=f(z+\overline{w}),\]
and Kobak considers the choice
\[f=2+\cos(z+\overline{z}+w+\overline{w})\]
as this gives an explicit, doubly-Hermitian metric on a torus $\mathbb{C}^2/2\pi\mathbb{Z}^4$. If we introduce real coordinates $x,y,u,v$ by
\[z=x+iy,\;\;\;w=u+iv\]
then we'll consider the more general class:
\[g=dx^2+dy^2+f(x+u)(du^2+dv^2).\]
We Wick-rotate by setting $y=it$ and change the sign to obtain the static Lorentzian metric
\be\label{dh1}g_L=dt^2-dx^2-f(x+u)(du^2+dv^2),\ee

We analyse this metric in the NP formalism by choosing the null tetrad
\[D=\frac{1}{\sqrt{2}}(\partial_t+\partial_x),\;\;\Delta=\frac{1}{\sqrt{2}}(\partial_t-\partial_x),\;\;\delta=\frac{1}{\sqrt{2}}F(\partial_u+i\partial_v),\;\;\overline{\delta}=\frac{1}{\sqrt{2}}F(\partial_u-i\partial_v),\]
with $F$ to be fixed, and
\[\ell=\frac{1}{\sqrt{2}}(dt-dx),\;\;n=\frac{1}{\sqrt{2}}(dt+dx),\;\;m=-\frac{1}{\sqrt{2}}fF(du+idv),\;\;\overline{m}=-\frac{1}{\sqrt{2}}fF(du-idv),\]
so that $f^2F^2=f$ and $F=f^{-1/2}$. We're restricting to metrics with $F(x+u)$. Now calculate the spin-coefficients to be
\[0=\kappa=\sigma=\nu=\lambda=\epsilon=\gamma=\tau=\pi,\]
and
\[\rho=\mu=\frac{F'}{\sqrt{2}F},\;\;\alpha=-\beta=\frac{F'}{2\sqrt{2}},\]
with $F'=F_x=F_u$. The curvature components are found to be
\[0=\Psi_0=\Psi_4=\Phi_{02},\;\;\Psi_1=-\Psi_3=-\Phi_{01}=\Phi_{21}=-\frac{F}{4}\left(\frac{F'}{F}\right)',\]
\[\Psi_2=\frac16\left(\frac{(1-F^2)F''}{F}+\frac{(F^2-1)(F')^2}{F^2}\right)=\frac{2(1-F^2)}{3F}\Psi_3,\]
\[\Phi_{11}=\frac14\left(FF''-\frac{(F^2+1)(F')^2}{F^2}\right),\]
\[\Lambda=\Phi_{11}+\Psi_2.\]
This will have invariants $I,J$ real but won't have $I^3/J^2$ constant (which is not a surprise as the metric is not vacuum).

There are two more gsf spinors which can be taken to be
\[\alpha^A=\iota^A+\frac{1}{F}o^A,\;\;\beta^A=\iota^A-Fo^A,\]
so the Weyl spinor and the Killing spinor if there is one are both proportional to
\[\chi_{ABCD}:=o_{(A}\iota_B\alpha_C\beta_{D)}=o_{(A}\iota_B\iota_C\iota_{D)}+\frac{(1-F^2)}{F}o_{(A}o_B\iota_C\iota_{D)}-o_{(A}o_Bo_C\iota_{D)}.\]
From the curvature components given above one sees that in fact
\[\psi_{ABCD}=4\Psi_3\chi_{ABCD},\]
with $\Psi_3$ real, and it can now be checked that
\[\omega_{ABCD}=F^{-1}\chi_{ABCD}\]
is a valence-4 Killing spinor. This follows by direct calculation, given the spin-coefficients above, but the result can also be obtained by solving the geodesic equation as we see next. Then $\psi_{ABCD}$ is proportional to the Killing spinor $\omega_{ABCD}$ with the real function of proportionality $4F\Psi_3$.

\subsubsection{Solving the geodesic equation}
Note we have three linearly independent Killing vectors which we can take to be
\[K_1=\partial_t=\frac{1}{\sqrt{2}}(D+\Delta),\;\;\;K_2=\partial_v=-\frac{i}{F\sqrt{2}}(\delta-\overline{\delta}),\]
\[K_3=\partial_x-\partial_u=\frac{1}{\sqrt{2}}(D-\Delta)-\frac{1}{F\sqrt{2}}(\delta+\overline{\delta}).\]
Consider the spinor field $po^A+q\iota^A$, then this is parallelly propagated along itself if
\[(po^A+q\iota^A)(\overline{p}o^{A'}+\overline{q}\iota^{A'})\nabla_{AA'}(po_B+q\iota_B)=0,\]
equivalently, $p,q$ satisfy the system
\[\dot{p}=p(\alpha(p\overline{q}-q\overline{p})-\rho q\overline{q}),\]
\[\dot{q}=q(-\alpha(p\overline{q}-q\overline{p})+\rho p\overline{p}).\]
We deduce at once that
\[|p|^2+|q|^2=\mbox{  constant},\]
which is the constant of motion associated with $K_1$. By constant rescaling of $p,q$ we can set this to one without loss of generality. Next notice that
\[\frac{{(pq)^.}}{pq}=\rho(|p|^2-|q|^2),\]
which is real, so that $(pq)/(\overline{p}\overline{q})$ is constant and by multiplying $p,q$ by a constant phase can be taken to be one without loss of generality. Now we can parametrise $p,q$ as
\[p=\cos\theta e^{i\phi},\;\;q=\sin\theta e^{-i\phi},\]
for some $\theta,\phi$.

For $\omega_{ABCD}$ to be a Killing spinor we need the following to be a constant of the motion for null geodesics:
\[\Omega:=\omega_{ABCD}(po^A+q\iota^A)(po^B+q\iota^B)(po^C+q\iota^C)(po^D+q\iota^D)=F^{-1}(pq^3+\frac{(1-F^2)}{F}p^2q^2-p^3q),\]
\[=F^{-1}(\sin^3\theta\cos\theta e^{-2i\phi}-\sin\theta\cos^3\theta e^{2i\phi}+\frac{(1-F^2)}{F}\sin^2\theta\cos^2\theta).\]
There are constants of the motion $c_i$ from the Killing vectors $K_i$ respectively, and these are given by
\[c_1=\frac{1}{\sqrt{2}}(|p|^2+|q|^2)=\frac{1}{\sqrt{2}},\;\;c_2=-\frac{i}{F\sqrt{2}}(p\overline{q}-q\overline{p})\]
and
\[c_3=\frac{1}{\sqrt{2}}(|p|^2-|q|^2)-\frac{1}{F\sqrt{2}}(p\overline{q}+q\overline{p}),\]
or in terms of $\theta,\phi$
\[c_2=\frac{\sqrt{2} \sin\theta\cos\theta\sin(2\phi)}{F},\;\;c_3=\frac{1}{\sqrt{2}}((\cos^2\theta-\sin^2\theta)-2F^{-1}\sin\theta\cos\theta\cos(2\phi)).\]
and now one calculates
\[\Omega=\frac12((c_2-ic_1)^2+c_3^2),\]
which is indeed a constant of the motion, as required.

Thus we do have a valence-4 Killing spinor, and the quantity $\Psi(s)$ arising in the discussion of the Penrose limit is $4\Psi_3 F\Omega$ so that $\Psi/\overline{\Psi}=\Omega/\overline{\Omega}$ which is constant along null geodesics: now all Penrose limits are diagonalisable. I believe this to be the first explicit example of an algebraically-general metric with a valence-4 Killing spinor, indecomposable in the sense of not being an outer product of lower valence Killing spinors -- as we've seen, there can be no vacuum examples.
\section*{Appendix: the Penrose limits of the Kasner solution}
This is an example of a diagonalisable 4-metric with at least some Penrose limits which are not diaonalisable. It is straightforward to compute as the geodesic equation is integrable; the vacuum Kasner metric is
\[g=dt^2-t^{2p}dx^2-t^{2q}dy^2-t^{2r}dz^2,\]
with real $p,q,r$ satisfying
\[p+q+r=1=p^2+q^2+r^2.\]
It has Killing vectors
\[K_1=\partial_x,\;\;K_2=\partial_y,\;\;K_3=\partial_z.\]
We choose the null tetrad
\[D=\frac{1}{\sqrt{2}}(\partial_t+t^{-p}\partial_x),\;\;\Delta=\frac{1}{\sqrt{2}}(\partial_t-t^{-p}\partial_x),\;\;\delta=\frac{1}{\sqrt{2}}(t^{-q}\partial_y+it^{-r}\partial_z),\]
when the nonzero spin coefficients are
\[\rho=-\mu=-\frac{(q+r)}{2\sqrt{2}t},\;\;\sigma=-\lambda=-\frac{(q-r)}{2\sqrt{2}t},\;\;\gamma=-\epsilon=-\frac{p}{2\sqrt{2}t},\]
and the nonzero Weyl spinor components are
\[\psi_0=\psi_4=\frac{p(q-r)}{2t^2},\;\;\psi_2=-\frac{qr}{2t^2}.\]
The Weyl spinor can be written as
\[\psi_{ABCD}=\psi_0(o_Ao_Bo_Co_D+\iota_A\iota_B\iota_C\iota_D)+6\psi_2o_{(A}o_B\iota_C\iota_{D)},\]
and is algebraically general provided $p,q,r$ are distinct, which we'll assume.

For the geodesics, suppose the spinor field $\alpha^A=Ao^A+B\iota^A$ is parallelly propagated in the sense
\[\alpha^A\overline{\alpha}^{A'}\nabla_{AA'}\alpha_B=0,\]
then this is equivalent to the pair of equations
\[\dot{A}=-\epsilon A(|A|^2-|B|^2)+B(\rho A\overline{B}+\sigma\overline{A}B),\]
\[\dot{B}=\epsilon B(|A|^2-|B|^2)+A(\rho \overline{A}B+\sigma A\overline{B}),\]
and there are three constants of the motion obtained from the Killing vectors:
\[c_1=\frac{1}{\sqrt{2}}t^p(|B|^2-|A|^2),\;\;c_2=-\frac{1}{\sqrt{2}}t^q(A\overline{B}+\overline{A}B),
\;\;c_3=-\frac{it^r}{\sqrt{2}}(-\overline{A}B+A\overline{B}),\]
with the dot being $d/ds$ and $\alpha^A\overline{\alpha}^{A'}\nabla_{AA'}s=1$. Given these three constants of integration, we obtain from the metric that
\[\dot{t}^2=c_1^2t^{-2p}+c_2^2t^{-2q}+c_3^2t^{-2r},\]
which establishes integrability of the geodesic equations, but note that we are asking for more: in the language of \cite{pr} the \emph{flag-plane} of the spinor $\alpha^A$ is parallelly-propagated, as well as the \emph{flag-pole}.

Having solved for $A,B$ and chosen a null geodesic, we take a Penrose limit for which we need
\[\Psi(s):=\psi_{ABCD}\alpha^A\alpha^B\alpha^C\alpha^D=\psi_0(A^4+B^4)+6\psi_2A^2B^2.\]
The Penrose limit is diagonalisable iff $\Sigma$ vanishes where
\[\Sigma=i(\overline{\Psi}\dot{\Psi}-\Psi\dot{\overline{\Psi}}).\]
It is a straightforward, if untidy, calculation to see that $\Sigma$ is a fifth-order polynomial in $A,B,\overline{A}$ and $\overline{B}$ with coefficients expressible in terms of the $c_i$ and functions of $t$, and that it is \emph{not} zero if the product $c_1c_2c_3$ is nonzero. Thus there are some diagonalisable Penrose limits when one or more $c_i$ vanishes but in general they are nondiagonalisable even though the Kasner metric itself is diagonal. Given the results of this paper this is now not a surprise as no algebraically-general vacuum solution has all its Penrose limits diagonalisable.

\end{document}